\documentclass[12pt]{article}
\usepackage{graphicx,psfrag,epsfig, color}
\usepackage{amssymb,amsmath,amscd,amsthm}
\usepackage{graphicx,psfrag,epsfig}

\usepackage{graphicx}
\usepackage[active]{srcltx}

\newtheorem{theorem}{Theorem}[section]

\newtheorem{lemma}[theorem]{Lemma}

\date{}

\begin{document}
 \author{ Evgeny Lakshtanov\thanks{Department of Mathematics, Aveiro University, Aveiro 3810, Portugal.  This work was supported by Portuguese funds through the CIDMA - Center for Research and Development in Mathematics and Applications and the Portuguese Foundation for Science and Technology (``FCT--Fund\c{c}\~{a}o para a Ci\^{e}ncia e a Tecnologia''), within project UID/MAT/0416/2013 (lakshtanov@ua.pt)}  \and
 Boris Vainberg\thanks{Department
of Mathematics and Statistics, University of North Carolina,
Charlotte, NC 28223, USA. The work was partially supported  by the NSF grant DMS-1410547 (brvainbe@uncc.edu).}}

\title{On reconstruction of complex-valued once differentiable conductivities}

\maketitle
\vspace{3mm}
 \hspace{2.7in}           {\it In memory of our dear friend}
                       {\it Yuri Safarov}
\vspace{4mm}

\begin{abstract}
The classical $\overline \partial$-method has been generalized recently \cite{lnv}, \cite{lnv2} to be used in the presence of exceptional points. We apply this generalization to solve Dirac inverse scattering problem with weak assumptions on smoothness of potentials. As a consequence, this provides an effective method of reconstruction of complex-valued one time differentiable conductivities in the inverse impedance tomography problem.
\end{abstract}


\section{Introduction}

Let $\mathcal O$ be a bounded connected domain in $\mathbb R^2$ with a smooth boundary.
The electrical impedance tomography problem (e.g., \cite{borcea}) concerns determining the impedance in the interior of $\mathcal O$, given simultaneous
measurements of direct or alternating electric currents and voltages at the boundary $\partial \mathcal O$.
If the magnetic permeability can be neglected, then the problem can be reduced to the inverse conductivity problem (ICP), which consists of reconstructing function $\gamma(z), z=(x,y) \in \mathcal O$, via the known, dense in some adequate topology, set of data $(u|_{\partial \mathcal O},\frac{\partial u}{\partial\nu}|_{\partial \mathcal O})$,  where
\begin{equation}\label{set27A}
\mbox{div}(\gamma \nabla u(z)) =0, ~  z\in \mathcal O.
\end{equation}
 Here $\nu$ is the unit outward normal to $\partial \mathcal O$, $\gamma(z) = \sigma(z)+ i \omega\epsilon(z)$, where $\sigma$ is the electric conductivity and $\epsilon$ is the electric permittivity. If the frequency $\omega$ is ignorably small, then one can assume that $\gamma$ is a real-valued function, otherwise it is supposed  to be a complex-valued function.

An extensive list of references on tomography problem can be found in review \cite{borcea}. Here we will mention only the papers that seem to be particularly related to the present work.
For real $\gamma$, the inverse conductivity problem has been reduced to the inverse problem for the Schr\"{o}dinger equation. The latter was solved by Nachman in \cite{nachman} in the class of twice differentiable conductivities. Later Brown and Uhlmann \cite{bu} (see also Francini \cite{fr} for the case of small Im$\gamma$) reduced ICP to the inverse problem for the Dirac equation, which has been solved in  \cite{bc1}, \cite{sung1}. This approach requires the existence of only one derivative of $\gamma$. The authors of \cite{bu} proved the uniqueness for ICP. Later Knudsen and Tamasan \cite {knud} extended this approach and obtained a method to reconstruct the conductivity. Finally, the ICP has been solved by Astala and Paivarinta in \cite{ap} for real conductivities when both $\gamma-1$ and $1/\gamma-1$ are in $L^\infty_{\rm comp}(\mathbb R^2)$.

If a complex conductivity has at least two derivatives, then one can reduce equation (\ref{set27A}) to the Schr\"{o}dinger equation and apply the reconstruction method of Bukhgeim \cite{bukh} or the one developed in our recent work \cite{lnv}. This approach does not work in the case of only one time differentiable conductivities. On the other hand, the ideas of \cite{bu}, \cite{fr} (and of related inverse scattering methods discussed in  \cite{bc1}, \cite{sung1}) are not applicable to general complex conductivities due to possible existence of the so called {\it exceptional points}.
In \cite{lnv} we generalized the $\overline \partial$-method for the Schr\"{o}dinger equation to the case when exceptional points exist. A prototype of this generalization was considered in section 8 of \cite{RN2}. In the current paper, we use the ideas from  \cite{lnv} to extend the $\overline \partial$-method to the inverse Dirac problem in the presence of exceptional points. This reduces the smoothness assumption on $\gamma$ in the ICP with complex conductivities.

Note that the $\overline \partial$-method for the inverse Dirac scattering problem is important per se as a tool for solving certain nonlinear equations. We believe that results presented below can be used for solving non-linear Davey-Stewartson and Ishimori  equations (see, e.g., \cite{bc2}) where the $\overline \partial$-method for the inverse Dirac problem has been used.

Thus we will split the ICP into two independent parts:

1. Solving the inverse Dirac scattering problem (sections 2-5).

2. Representing the scattering data (\ref{27dec2}) of the Dirac scattering problem via the D-t-N map for equation (\ref{set27A}) (section 6).

The first problem will be solved (Theorem \ref{mthm}) if $ \gamma-1 \in W^{1,p}_{\rm{comp}}(\mathbb R^2), p>4, $ and $\mathcal F(\nabla \gamma )\in L^{2-\varepsilon}(\mathbb R^2)$ (here $\mathcal F$ is the Fourier transform). The second problem will be solved under a much less restrictive condition that $\gamma-1 \in L^{\infty}_{\rm comp}(\mathbb R^2)$. Note that solutions of the second problem can be found in \cite{knud} (with three equations for two unknown functions) and \cite{hs} (in the case of absence of exceptional points). We will closely follow the approach used in \cite{hs} (section 3) combined with some ideas from \cite{knud} to extend the results of \cite{hs} to the situation when exceptional points are present.

Our smoothness requirement for the reconstruction of the complex conductivities is close to the one used in the case of real-valued conductivities. Indeed, while $\gamma -1 \in W^{1,p}_{\rm comp}(\mathbb R^2), p>2, $ is assumed in order to prove a uniqueness result \cite{bu} in the class of real-valued conductivities, additional smoothness is usually required for the reconstruction of the conductivities. For example, $\gamma\in C^{1+\varepsilon}$ is assumed in \cite{knud}. The difference in assumptions for the uniqueness and reconstruction results was also mentioned in  \cite{afr}, \cite{blasten}, \cite{ns}. Similarly, our assumption $\mathcal F(\nabla \gamma )\in L^{2-\varepsilon}(\mathbb R^2)$ requires $\gamma$ to be a little bit smoother than just $ \gamma-1 \in W^{1,p}_{\rm{comp}}(\mathbb R^2), p>4$.

Below $z$ will be considered as a point of a complex plane, i.e.,
 $$
 z=x+iy\in\mathbb C,
 $$
 and $\mathcal O$ will be considered as a domain in $\mathbb C$.

The following observation made in \cite{bu}, \cite{fr} plays an important role. Let $u$ be a solution of (\ref{set27A}) and $\partial = \frac 1 2 \left (\frac{\partial}{\partial x} - i\frac{\partial}{\partial y} \right )$. Then the pair
$\phi=\gamma^{1/2}(\partial u, \overline{\partial} u)^t=\gamma^{1/2}\left(
\!\!\!\begin{array}{c}
                                                                                            \partial u \\
                                                                                          \overline{\partial} u\\
                                                                                         \end{array} \!\!\!
                                                                                       \right)$
satisfies the Dirac equation
\begin{equation}\label{firbc}
\left ( \begin{array}{cc} \overline{\partial} & 0 \\ 0  & \partial \end{array} \right ) \phi = q {\phi}, \quad z=x+iy  \in \mathcal O,
\end{equation}
where
\begin{eqnarray}\label{char1bc}
q(z)=\left ( \begin{array}{cc}0 &q_{12}(z) \\
q_{21}(z) &  0\end{array} \right ), \quad q_{12}=-\frac{1}{2}\partial \log \gamma, \quad q_{21}=-\frac{1}{2}\overline{\partial }\log \gamma.
\end{eqnarray}
Thus it is enough to solve
the inverse Dirac scattering problem instead of the ICP. If it is solved and $q$ is found, then conductivity $\gamma$ can be immediately found from (\ref{char1bc}).

In order to complete the reduction of the ICP to the inverse Dirac problem, one needs only to show that the scattering data for the Dirac equation can be found via $\left (u\!\left |_{\partial \mathcal O}\right .,\frac{\partial u}{\partial\nu}\!\left |_{\partial \mathcal O}\right. \right)$. In fact, the scattering data for the Dirac equation can be obtained by simple integration of the Dirichlet data for the same equation, see formula (\ref{2106A}). In the last section we will show that the latter data can be found via known
$\left (u\!\left |_{\partial \mathcal O}\right .,\frac{\partial u}{\partial\nu}\! \left |_{\partial \mathcal O}\right. \right)$. This will complete the reconstruction of conductivity.

 We assume  that $\log \gamma$ is well defined in the whole complex plane, e.g., there exist a ray that does not intersect the range of $\gamma$.

\section{Main results}
We will use a different form of equation (\ref{firbc}) in sections 2-5: instead of Beals-Coifmann notations $\phi=(\phi_1,\phi_2)^t$, we will rewrite the equation in Sung notations: $\psi_1=\phi_1,\psi_2=\overline{\phi_2}$. Then the {\it vector} $\psi=(\psi_1,\psi_2)^t$ is a solution of the following system
\begin{equation}\label{fir}
\overline{\partial }\psi = Q \overline{\psi},
\end{equation}
where
\begin{eqnarray}\label{char1}
Q(z)=\left ( \begin{array}{cc}0 &Q_{12}(z) \\
Q_{21}(z) &  0\end{array} \right ), \quad q_{12}=Q_{12}, \quad Q_{21}=\overline{q_{21}}.
\end{eqnarray}

 Consider the {\it matrix} solution of (\ref{fir}) that depends on parameter $k \in \mathbb C$ and has the following behavior at infinity:
\begin{equation}\label{lim}
\psi(z,k) e^{-i\overline{k}z/2} \rightarrow I, ~ z \rightarrow \infty.
\end{equation}
Note that the plane waves
\begin{equation}\label{exp}
\varphi_0(k,z):=e^{i\overline{k}z/2}, \quad k \in \mathbb C,
\end{equation}
are growing at infinity exponentially in some directions, and the same is true for the elements of the matrix $\psi(z,k)$.

Problem (\ref{fir})-(\ref{lim}) is equivalent (e.g., \cite{sung1}) to the
Lippmann-Schwinger equation:

\begin{eqnarray}\label{19JanA}
\psi(z,k)=e^{i\overline{k}z/2}I+ \int_{\mathbb C} G(z-z',k) Q(z') \overline{\psi}(z',k) dx' dy', \\
G(z,k)=\frac{1}{\pi}\frac{e^{i\overline{k}z/2 }}{z}\label{2DecA}.
\end{eqnarray}

Denote
\begin{equation}\label{mu1}
\mu(z,k)= \psi(z,k)e^{-i\overline{k}z/2}= \psi(z,k) \varphi_0(k,-z).
\end{equation}

Let $Q_{12},Q_{12}\in L^p_{\rm{comp}}(\mathbb R^2), ~p>2$. Here and below we use the same notations for functional spaces, irrespectively of whether those are the spaces of matrix or scalar valued functions. Let us make the substitution $\psi \to\mu(\cdot,k) -I$ in (\ref{19JanA}). It was proved in \cite[Th.A. iii]{bu} that equation (\ref{19JanA}) after the substitution becomes Fredholm in $L^q(\mathbb R^2), ~ q>2p/(p-2)$.
 Solutions $\psi$ of (\ref{19JanA}) are called the {\it scattering solutions}, and the values of $k$ such that the homogeneous equation (\ref{19JanA}) has a non-trivial solution are called {\it exceptional points}. The set of exceptional points will be denoted by $\mathcal E$. Thus the scattering solution may not exist if $k\in \mathcal E$.

Let us define a matrix that is called the {\it (generalized) scattering data}. It is given by the formula
\begin{equation}\label{27dec2}
h(\varsigma,k)  =  \frac{1}{(2\pi)^2} \int_{\mathbb C}  \varphi_0(-z,\varsigma)Q(z)\overline{\psi}(z,k)dxdy, ~ \varsigma \in \mathbb C,~ k \not \in \mathcal E,
\end{equation}
which can be rewritten as
\begin{equation}\label{14Abr1}
{h}(\varsigma,k)  =  \frac{1}{(2\pi)^2} \int_{\mathbb C}  e^{-i(k \overline{z} +\overline{\varsigma }z)/2} Q(z)\overline{\mu}(z,k)dxdy, ~ \varsigma \in \mathbb C,~ k \not \in \mathcal E.
\end{equation}
To justify the use of the term {\it scattering data} we can use Green formula for a regular domain $\mathcal O$ and function $f$:
 $$
\int_{\partial \mathcal O} fdz = 2i \int_{\mathcal O} \overline \partial {f} dx dy.
$$
Then $h$ can be rewritten as follows:
\begin{equation}\label{2106A}
  h(\varsigma,k)  =\frac{1}{(2\pi)^2} \int_{\partial\mathcal O}\varphi_0(-z,\varsigma)\overline{\psi}(z,k)dz, ~ \varsigma \in \mathbb C,~ k \not \in \mathcal E,
\end{equation}
where $\nu $ is the outer unit normal to $\partial\mathcal O$. Thus, one does not need to know the potential $Q$ in order to find the values of $h$. The latter  matrix can be evaluated if the Dirichlet data $\psi|_{\partial\mathcal O}$ is known for equation (\ref{fir}).

There are no exceptional points in some neighborhood of $k=\infty$ (e.g., \cite[lemma 2.8]{sung1}, \cite[lemma C]{bu}) and for each $\varepsilon >0$ there exists some $R(\varepsilon)$ such that
\begin{equation}\label{1006A}
\|\mu-I\| < \varepsilon \mbox{ in } L^\infty_z(L^\infty_k(\{k:|k|>R(\varepsilon)\}))
\end{equation}

Let us choose $A$ large enough and $k_0 \in \mathbb C$  such that all the exceptional points for both conductivities $\gamma$ and $1/\gamma$ (i.e., for potentials $Q(\cdot)$ and $-Q(\cdot)$) are contained in the disk
\begin{equation}\label{set28A}
D=\{  k \in \mathbb C: 0 \leq |k| < A\},
\end{equation}
and that $k_0 $ belongs to the same disc $ D$ and is not exceptional for both conductivities $\gamma$ and $1/\gamma$.

 Consider the space
\[
\mathcal H^s = \left \{ u \in L^s(\mathbb R^2_k)\bigcap C(D) \right   \}, \quad s>2.
\]
Recall that we use the same notation for matrices if their entries belong to $\mathcal H^s$.
 Let $T_z:\mathcal H^s \to \mathcal H^s$ be the operator defined by the formula
\begin{eqnarray}\nonumber
T_z \phi (k)=
\frac{1}{\pi} \int_{\mathbb C\backslash D}  e^{i(\overline{\varsigma}z+\overline{z}{\varsigma})/2} \overline{\phi}(\varsigma) h^o(\varsigma,\varsigma)  \frac{d\varsigma_R d\varsigma_I}{\varsigma-k} + \\\label{glavOperator}
\frac{1}{2\pi i} \int_{\partial D} \frac{d\varsigma}{\varsigma - k} \int_{\partial D}
\left [e^{i/2(\varsigma\overline{z}+\overline{\varsigma'}z)}\overline{\phi^-(\varsigma')} \Pi^o  + e^{i/2(\varsigma-{\varsigma'})\overline{z}}\phi^-(\varsigma')\Pi^d \mathcal C \right ]  \! \!\!\left [{\rm Ln}\frac{\overline{\varsigma'}-\overline{\varsigma} }{\overline{\varsigma'}-\overline{k_0} }h(\varsigma',\varsigma) d\overline{\varsigma'} \right ]\!,
\end{eqnarray}
where $ \phi\in \mathcal H^s$,
$\phi^-$ is the boundary trace of $\phi$ from the interior of $D$,
$\mathcal C$ is the operator of complex conjugation, $\Pi^oM=M^o$ is the off-diagonal part of a matrix $M$, $\Pi^dM=M^d$ is the diagonal part, and $h^o=\Pi^o h$. The logarithmic function here is well defined, see the Remark after Lemma \ref{1DecLA}.


We'll use the word {\it generic} when referring to elements that belong to an open and dense subset $V$ of a topological space $S$.

Let $S_{\varepsilon,p}$ be the space of potentials $Q$ with support in $\mathcal O$ such that $Q \in L^{p}_{\rm{comp}}(\mathbb R^2), p>4,$ and $\mathcal FQ \in L^{2-\varepsilon}(\mathbb R^2), ~ \varepsilon>0$.
\begin{theorem}\label{mthm} Let $Q_{12},Q_{21}\in S_{\varepsilon,p}$.   Then for each  $\widetilde{s}>\max(\frac{2p}{p-4},\frac{4}{\varepsilon}-2)$ the following statements are valid.
\begin{itemize}
\item
Operator $T_z$ is compact in $\mathcal H^{\widetilde s}$ for all $z \in \mathbb C$ and depends continuously on $z$.
  \item  Let us fix $z_0 \in \mathbb C$. Then for generic potentials $Q_{12},Q_{21}$ in $S_{\varepsilon,p}$, equation
  \begin{equation}\label{inteq}
 (I+T_z)w_z =-T_z I
 \end{equation}
  is uniquely solvable in $\mathcal H^{\widetilde s}$ for all $z$ in some neighborhood of $z_0$ (the neighborhood may depend on $Q$).
  \item For  $k \not \in \mathcal E$,  function $\psi=[e^{i\overline{k}{z}/2} \mathcal C \Pi^d + e^{-i\overline{z}k/2} \Pi^o](w_z+I)$, where $w_z(\cdot)$ is the solution of (\ref{inteq}),
satisfies the equation $ \overline{\partial}\psi = Q \overline{\psi}$ in $\mathcal O$.
   \end{itemize}
\end{theorem}
\noindent
{\bf Remarks.} 1) After equation (\ref{inteq}) is solved and $\psi$ defined in the last item of the theorem is found, one can immediately reconstruct potential $Q$ from (\ref{fir}): $Q=\frac{\partial \psi}{\partial \overline{z}} \overline{\psi}^{~\!-1}, |k|\gg 1.$ Note that det$\psi \neq 0$ for large $|k|$ due to (\ref{1006A}).

2) Consider a set of conductivities $\gamma^a$ that depend on power $a \in (0,1]$. These conductivities correspond to potentials $aQ,~ a \in (0,1]$ (see (\ref{char1})). Then item 2 of the theorem can be replaced by the following statement: equation (\ref{inteq}) with $Q$ replaced by $a Q_0$ with a fixed potential $Q_0$ is uniquely solvable for a set of parameters $(z, a)\in\mathcal O \times (0,1]$ of full 3D-measure.  In fact, it will be proved that, for each $z$, the unique solvability can be violated for at most finitely many values of $a=a_j(z),~z\in \mathcal O$.

3) If the kernel $h^o(\varsigma,\varsigma)$ is truncated (as it is usually done in numeric applications), then operator $T_z$ becomes analytic in $\Re z, \Im z$, and therefore the invertibility of (\ref{inteq}) at a point $z_0$ implies its invertibility at a.e. point of $\mathbb C$.

\section{Derivation of the integral equation}
Following  \cite{sung1}, we will work with the matrix
\begin{equation}\label{vvv}
v=\left ( \begin{array}{cc} \overline{\mu_{11}}(z,k) & \mu_{12}(z,k)e^{i(\overline{k}z+\overline{z}k)/2} \\
\mu_{21}(z,k)e^{i(\overline{k}z+\overline{z}k)/2} &  \overline{\mu_{22}}(z,k)\end{array} \right ), \quad k\in \mathbb C \backslash D,
\end{equation}
instead of $\mu$. It was shown in \cite{sung1} that
\begin{equation}\label{dbar}
\frac{\partial}{\partial \overline{k}} v(z,k)= e^{i(\overline{k}z+\overline{z}k)/2} \overline{v}(z,k) h^o(k,k) =: \mathcal Tv, \quad k\in \mathbb C \backslash D,
\end{equation}
 where $h^o=\Pi^o h$.

We introduce the matrix function $\psi^+(z,k)=\psi^+(z,k,k_0)$ as the solution of the Lippmann-Schwinger equation
\begin{equation}\label{eq1}
\psi^+(z,k) =e^{i\overline{k}z/2}I  +
\int_{ \mathbb C}  G(z-\varsigma,k_0) Q(\varsigma) \overline{\psi^+}(\varsigma,k)d\varsigma_R d\varsigma_I, \quad  k\in \overline{D},
\end{equation}
where $k_0$ was introduced in (\ref{set28A}). This equation is similar to (\ref{19JanA}), but the parameter $k=k_0$ in the argument of $G$ is fixed now. We define
\begin{equation}\label{mupl}
\mu^+(z,k)=\{\mu^+_{ij}(z,k)\}:=\psi^+ e^{-i\overline{k}z/2}, \quad  k\in \overline{D},
\end{equation} and $v^+$ as
\begin{equation}\label{1504A}
v^+=\left ( \begin{array}{cc} \overline{\mu^+_{11}}(z,k) & \mu^+_{12}(z,k)e^{i(\overline{k}z+\overline{z}k)/2} \\
\mu^+_{21}(z,k)e^{i(\overline{k}z+\overline{z}k)/2} &  \overline{\mu^+_{22}}(z,k)\end{array} \right ), \quad  k\in \overline{D}.
\end{equation}

\begin{lemma}\label{15J1} The following relation holds
\begin{equation}\label{1504B}
\frac{\partial v^+}{\partial \overline{k}} =0, \quad  k\in D.
\end{equation}
\end{lemma}
\noindent
{\bf Proof.}\label{15apr}
Denote $\mu^+_{0}=\{\mu^+_{ij,0}\}:=\psi^+ e^{-i\overline{k_0}z/2}$. Thus matrix $\mu^+_{0}$ has the same form as matrix $\mu^+$ defined by (\ref{mupl}), but now the value of $k$ in the exponent is fixed. In other words,
\begin{equation}\label{mu0}
\mu^+=\mu^+_0e^{i(\overline{k_0}-\overline{k})z/2 }.
\end{equation}
Let
\begin{equation}\label{muSet5}
\mathcal L_k \varphi (z)= \frac{1}{\pi}\int_{\mathbb C} \varphi(w)\frac{e^{-i\Re(\overline{k}w)} dw_R dw_I}{z-w}.
\end{equation}
Then equation (\ref{eq1}) implies that
\begin{eqnarray}
\mu^+_{11,0}=e^{i(\overline{k}-\overline{k}_0)z/2} + \mathcal L_{k_0} [Q_{12}(z') \overline{\mu_{21,0}^+}(z',k)],
\\
\mu^+_{21,0}= \mathcal L_{k_0} [Q_{21}(z') \overline{\mu^+_{11,0}}(z',k)],
\end{eqnarray}
and therefore
\begin{eqnarray}
\mu^+_{11,0}=e^{i(\overline{k}-\overline{k}_0)z/2} + \mathcal L_{k_0} [Q_{12}(z') \overline{\mathcal L_{k_0}} [\overline{Q_{21}}(z') {\mu^+_{11,0}}(z',k)]],
\\
\mu^+_{21,0}= \mathcal L_{k_0} \left [ Q_{21}(z') \left ( e^{i({k}-{k}_0)\overline{z}/2} + \overline{\mathcal L_{k_0}} [\overline{Q_{12}}(z') {\mu^+_{21,0}}(z',k)] \right ) \right ].
\end{eqnarray}
These equations are Fredholm with empty kernel due to the choice of the point $k_0$ since operator $I-(G_{k_0} Q\mathcal C)^2$ is invertible if both operators $I-G_{k_0}Q\mathcal C$ and $I+G_{k_0} Q\mathcal C$ are invertible. Here $G_{k_0}$  is the convolution operator with the kernel $G(z,k_0)$ and $\mathcal C$ is the operator of complex conjugation. Thus $\mu^+_{11,0}$ is analytic in $\overline{k}$ and $\mu^+_{21,0}$ is analytic in $k$, i.e.,
$$
\frac{\partial \mu^+_{11,0}}{\partial {k}} =0, \quad \frac{\partial \mu^+_{21,0}}{\partial \overline{k}} =0.
$$
Using (\ref{mu0})
we get
\begin{eqnarray}\label{19Apr3}
\frac{\partial v^+_{11}}{\partial \overline{k}} = {\frac{\partial \overline{\mu^+_{11}}}{\partial \overline{k}} } ={\frac{\partial \overline{\mu^+_{11,0} e^{i(\overline{k_0}-\overline{k})z/2 }}}{\partial \overline{k}} }=0,
\\ \nonumber
 \frac{\partial v^+_{21}}{{\partial \overline{k}}} = \frac{\partial \mu^+_{21,0}e^{i(\overline{k_0}-\overline{k})z/2 }e^{i\Re(\overline{k}z)}}{{\partial \overline{k}}} =\frac{\partial \mu^+_{21,0}e^{i(\overline{k_0}z+\overline{z}k)/2 }}{{\partial \overline{k}}}= 0.
\end{eqnarray}
One can prove similarly that
\begin{equation}\label{19Apr3}
\frac{\partial v^+_{22}}{\partial \overline{k}} = 0, \quad
\frac{\partial v^+_{12}}{\partial \overline{k}} = 0.
\end{equation}
\qed

For each $z \in \mathbb C$, consider the following matrix function $v'$:
\begin{equation}\label{pspr}
v'(z,k) := \left \{ \begin{array}{l}
v(z,k), \quad k\in \mathbb C \backslash D, \\
v^+(z,k), \quad  k\in D.
\end{array}
\right .
\end{equation}
From (\ref{dbar}) and Lemma \ref{15J1} it follows that
\begin{equation}\label{15Decdbar}
\frac{\partial}{\partial \overline{k}} v' (z,k)=\mathcal T'(k) v' (z,\cdot), \quad k \notin \partial D,
\end{equation}
where the operator $\mathcal T'$ is given by
\begin{equation}\label{1504C}
\mathcal T' =  \left \{ \begin{array}{l}
\mathcal T(k), \quad k\in \mathbb C \backslash D, \\
0, \quad  k\in D,
\end{array} \right .
\end{equation}
and $\mathcal T$ is defined in (\ref{dbar}).

Our main goal in this section is to prove the following statement.

\begin{lemma}\label{0210L1}
If $\psi$ satisfies the Lippmann-Schwinger equation (\ref{19JanA}), then the matrix function $w=v'-I$ with $v'$ defined by (\ref{pspr}) satisfies equation (\ref{inteq}), where $T_z$ is given by (\ref{glavOperator}).
\end{lemma}
Before we proceed with the proof of this lemma, we need to describe the boundary condition for the matrix function $v'$ at $\partial D$. We start by noting that (\ref{vvv}), (\ref{1504A}) together with (\ref{mu1}) and (\ref{mupl})
 imply that
\begin{eqnarray}\nonumber
\Pi^d(v-v^+) =\overline{\Pi^d (\psi-\psi^+)}e^{ik\overline{z}/2}, \quad \Pi^o(v-v^+) =  \Pi^o(\psi-\psi^+) e^{-i\overline{k}z/2} e^{i\Re{(\overline{k}z)}},
\\\label{20Apr1}
\Pi^d\psi^+ =\overline{\Pi^d v^+} e^{i\overline{k}{z}/2}, \quad \Pi^o\psi^+ =  \Pi^ov^+ e^{i\overline{k}z/2} e^{-i\Re{(\overline{k}z)}}.
\end{eqnarray}

The latter relations can be rewritten as follows:
\begin{eqnarray}\label{20Apr2}
v-v^+=\mathcal A(k,z) (\psi-\psi^+), \quad
\psi^+ = \mathcal A^*(k,z)v^+,
\end{eqnarray}
where
\begin{equation}\label{19Apr2}
\mathcal A(k,z)=[e^{ik\overline{z}/2}\mathcal C \Pi^d + e^{-i\overline{k}z/2} e^{i\Re{\overline{k}z}}\Pi^o], \quad \mathcal A^*(k,z) = [e^{i\overline{k}{z}/2} \mathcal C \Pi^d + e^{i\overline{k}z/2} e^{-i\Re{\overline{k}z}} \Pi^o].
\end{equation}

\begin{lemma}\label{lemma12NovA} The following integral equation holds for each $z \in \mathbb C$
\begin{equation}\label{11NovC}
v'(z,k)-I = \frac{-1}{\pi} \int_{\mathbb C }  (\mathcal T'v')(\varsigma) \frac{d\varsigma_R d\varsigma_I}{\varsigma-k} + \frac{1}{2\pi i} \int_{\partial D} \frac{[v'](z,\varsigma)}{\varsigma - k}d\varsigma,
\end{equation}
where $[v']=v^+-v$ is the jump of $v'$ on $\partial D$.
\end{lemma}
\noindent
{\bf Remark.} Here and everywhere below, the direction of integration over the boundary of a domain is chosen in such a way that the domain remains on the left during the motion along of the boundary.
\\
{\bf Proof.} The following Cauchy-Green formulas hold for each $f\in C^1(\overline \Omega)$ and an arbitrary bounded domain $\Omega$ with a smooth boundary:
\begin{eqnarray}\label{2DecB}
f(k)= - \frac{1}{\pi} \int_{ \Omega} \frac{\partial f (\varsigma)}{\partial \overline \varsigma} \frac{d\varsigma_R d\varsigma_I}{\varsigma-k} + \frac{1}{2\pi i} \int_{\partial   \Omega} \frac{f(\varsigma)}{\varsigma - k}d\varsigma, \quad k \in  \Omega, \\ \label{2DecC}
0= - \frac{1}{\pi} \int_{ \Omega} \frac{\partial f (\varsigma)}{\partial \overline \varsigma} \frac{d\varsigma_R d\varsigma_I}{\varsigma-k} + \frac{1}{2\pi i} \int_{\partial \Omega} \frac{f(\varsigma)}{\varsigma - k}d\varsigma,\quad k \not \in \overline{ \Omega}.
\end{eqnarray}

Denote by $D_R$ the disk $D$ with the constant $A$ replaced by $R>A$, i.e., $D_R=\{k\in \mathbb C: |k| < R\}$. Let $D_R^-=D_R\backslash D $. Assume that $k\in D_R^-$. We add the left- and right-hand sides in formulas (\ref{2DecB}) and (\ref{2DecC}) with $f=v'$ in both formulas and $\Omega=D_R^-$ in (\ref{2DecB}) and $\Omega=D$ in (\ref{2DecC}). If we take (\ref{15Decdbar}) into account, we obtain that
\begin{equation}\label{11NovA}
v'(z,k) = -\frac{1}{\pi} \int_{D^-_R}  (\mathcal T'v')(z,\varsigma) \frac{d\varsigma_R d\varsigma_I}{\varsigma-k} + \frac{1}{2\pi i} \int_{\partial D} \frac{[v']}{\varsigma - k}d\varsigma+ \frac{1}{2\pi i} \int_{\partial D_R} \frac{v'}{\varsigma - k}d\varsigma.
\end{equation}
It remains to note that the last term on the right converges to the unit matrix as $R\to\infty$,  due to (\ref{1006A}).

\qed

Equation (\ref{11NovC}) does not take into account the fact that the matrix functions $\psi$ and $\psi^+$ are related. Our next goal is to take that relation into account and change the last term in (\ref{11NovC}). The first step in this direction is
\begin{lemma}\label{1DecLA}
Let
\begin{equation}\label{20Apr3}
W(k,\varsigma): =  {\rm Ln}\frac{\overline{\varsigma}-\overline{k} }{\overline{\varsigma}-\overline{k_0} }, \quad~ k,\varsigma \in \partial D,
\end{equation}
where $k_0$ was introduced in (\ref{set28A}). Then the following relation holds
\begin{equation}\label{13NovB}
G(z,k)-G(z,k_0) =\frac{1}{(2\pi)^2} \int_{\partial D} W(k,\varsigma) e^{i\overline{\varsigma} z/2}d\overline{\varsigma}, \quad~ k \in \partial D.
\end{equation}
\end{lemma}
\noindent
{\bf Remark.}
Let us move the origin in $\mathbb C$ into the point $\overline{\zeta}$, and then make the rotation of axis such that the direction of the $x$-axis is defined by the vector from $\overline{\zeta}$ to $-\overline{\zeta}$.
Then $|\arg(\overline{\varsigma}-\overline{k})|\leq \pi/2$ and $|\arg(\overline{\varsigma}-\overline{k_0})|< \pi/2$, i.e.,
\begin{equation}\label{arg}
\left |\arg \frac{\overline{\varsigma}-\overline{k} }{\overline{\varsigma}-\overline{k_0} } \right | <\pi, \quad \varsigma,k \in \partial D.
\end{equation}
Hence, function (\ref{20Apr3}) is well defined.
\\
{\bf Proof.}
We apply the Cauchy formula to (\ref{2DecA}) and obtain that
$$
\frac{\partial}{\partial\overline{k}} G(z,k) = \frac{1}{(2\pi)^2}
\int_{\partial D} \frac{d\overline{\varsigma}}{\overline{\varsigma}- \overline{k}} e^{i\overline{\varsigma}z/2}, \quad k \in D.
$$
From  (\ref{2DecA}) it also follows that
\begin{equation*}\label{2DecF1}
\frac{\partial}{\partial {k}} G(z,k) = 0, \quad k \in D.
 \end{equation*}
We reconstruct $G$ from its gradient and obtain:
$$
G(z,k)-G(z,k_0) = \int_{k_0}^k \frac{\partial}{\partial {\overline k}} G(z,k) d\overline{k} =  \frac{1}{(2\pi)^2}\int_{k_0}^k \int_{\partial D} \frac{d\overline{\varsigma}}{\overline{\varsigma}- \overline{k}} e^{i\overline{\varsigma}z/2}d\overline k.
$$
It remains only to change the order of integration.

\qed

Now we are in a position to express $\psi-\psi^+$ in (\ref{11NovC}) in the form of a compact operator applied to $\psi^+$.
\begin{lemma}\label{13NovA}
The following representation holds
\begin{equation}\label{1DecC}
\psi(z,k)= \psi^+(z,k) + \int_{\partial D} (\Pi^d \psi^+(z,\varsigma) + \Pi^o\psi^+(z,\varsigma) \mathcal C) [W(k,\varsigma)h(\varsigma,k)d\overline{\varsigma}], \quad k\in \partial D,
\end{equation}
where $W(k,\varsigma)$ is given by formula (\ref{20Apr3}) and $h$ is defined in (\ref{27dec2}).
\end{lemma}
\noindent
{\bf Proof.}
Recall that $\varphi_0(z,k) = e^{i\overline{k}z/2}$. We will denote by $G_{k_0}, G_k$  the convolution operators with the kernels $G(z,k_0), G(z,k)$.
Then one can rewrite (\ref{eq1}) and (\ref{19JanA}) as follows
\begin{equation}\label{prp1}
\psi^+(z,k)= (I-G_{k_0}Q\mathcal C)^{-1} \varphi_0(z,k)I, \quad
\psi(z,k)= (I-G_kQ \mathcal C)^{-1} \varphi_0(z,k)I.
\end{equation}
Thus
$$
\psi^+(z,k)= (I-G_{k_0}Q\mathcal C)^{-1} [(I-G_k Q \mathcal C)\psi(z,k)],
$$
and therefore
\begin{equation}\label{prp}
\psi(z,k)-\psi^+(z,k) = (I-G_{k_0}Q \mathcal C)^{-1} (G_k - G_{k_0})(Q(\cdot) \mathcal C\psi(\cdot,k) ).
\end{equation}

We evaluate $G_k - G_{k_0}$ using Lemma \ref{1DecLA} and the obvious relation that $\varphi_0(z-u,k)=\varphi_0(z,k)\varphi_0(-u,k)$. This leads to
\begin{eqnarray*}
(G_k - G_{k_0})(Q\mathcal C  \psi(\cdot,k) )\\ = \frac{1}{(2\pi)^2}  \int_{\partial D} W(k,\varsigma) \varphi_0(z,\varsigma)\left (\int_{\mathbb C}\varphi_0(-u,\varsigma)Q(u) \overline{\psi}(u,k)du_Idu_R \right ) d\overline{\varsigma} \\=  \int_{\partial D} W(k,\varsigma) \varphi_0(z,\varsigma)h(\varsigma,k)d\overline{\varsigma} .
\end{eqnarray*}
We plug the last relation into (\ref{prp}). Note that operator $(I-G_{k_0}Q\mathcal C)$ contains factor $\mathcal C$, and therefore it is nonlinear with respect to multiplication by complex numbers. Formula (\ref{prp1}) implies that $(I-G_{k_0}Q\mathcal C)^{-1}\varphi_0(\cdot,\varsigma)I =\psi^+(z,\varsigma)$ and
$$
(I-G_{k_0}Q\mathcal C)^{-1}(i\varphi_0(\cdot,\varsigma)I) =i(\Pi^d - \Pi^o)\psi^+(z,\varsigma).
$$
The validity of the last equality is easy to verify directly if the matrix equation (\ref{eq1}) is written in a component-wise fashion.
Formula (\ref{prp}) and the last two relations imply (\ref{1DecC}).
\qed
\\
 {\bf Proof of lemma \ref{0210L1}.}
 We put (\ref{1DecC}) into (\ref{11NovC}) and obtain equation (\ref{inteq}) with
\begin{eqnarray}\nonumber
T_z \phi (k)=
\frac{1}{\pi} \int_{\mathbb C \backslash D}  e^{i(\overline{\varsigma}z+\overline{z}{\varsigma})/2} \overline{\phi}(\varsigma) h^o(\varsigma,\varsigma)  \frac{d\varsigma_R d\varsigma_I}{\varsigma-k} + \\
\frac{1}{2\pi i} \int_{\partial D} \frac{d\varsigma}{\varsigma - k} \int_{\partial D}
\mathcal A(\varsigma,z) \mathcal [\Pi^d A^*(\varsigma',z) \phi^-(\varsigma') + \Pi^o A^*(\varsigma',z) \phi^-(\varsigma')\mathcal C] \left [{\rm Ln}\frac{\overline{\varsigma'}-\overline{\varsigma} }{\overline{\varsigma'}-\overline{k_0} }h(\varsigma',\varsigma) d\overline{\varsigma} \right ].
\end{eqnarray}
Then an explicit straightforward calculation shows that $T_z$  can be simplified to (\ref{glavOperator}).

 \qed

\section{Analysis of scattering data and of operator $T_z$}
\begin{lemma}\label{lonE}
If $Q_{12},Q_{21} \in L^p_{\rm comp}(\mathbb R^2), ~p>4$, and $\mathcal F Q_{12}, \mathcal FQ_{21} \in L^{2-\varepsilon}(\mathbb R^2)$, then $h_{12}(k,k)$, $h_{21}(k,k) \in L^s(\mathbb R^2\backslash D)$ for each $s>\max\left (\frac{p}{p-2},2\!-\!\varepsilon \right )$.
\end{lemma}
\noindent
{\bf Proof}.
From (\ref{19JanA}),(\ref{mu1}),(\ref{muSet5})
\begin{equation}\label{2006A}
\mu=I+ \mathcal L_k Q_{} (I+\overline{\mathcal L_k} \overline{Q}_{} )\mu.
\end{equation}
Thus
\begin{equation}\label{2006C}
\overline{\mu_{11}}=1+
\frac{1}{\pi^2} \int_{\mathbb C} d{S_1} \int_{\mathbb C} dS_2 \frac{e^{i \Re(k\overline{z}_1)}}{\overline{z}-\overline{z_1}} \overline{Q}_{12}(z_1) \frac{e^{-i \Re(k\overline{z}_2)}}{z_1-z_2} {Q_{21}}(z_2) \overline{\mu_{11}}(z_2,k),
\end{equation}
where $dS=dz_R d z_I$.
Recall that
\begin{equation}\label{charl2}
{h_{21}}(k,k)  =  \frac{1}{(2\pi)^2} \int_{\mathbb C}  e^{-i\Re(k\overline{z})} Q_{21}(z)\overline{\mu}_{11}(z,k)dS.
\end{equation}
We replace $\overline{\mu}_{11}$  in (\ref{charl2}) by the right hand side of (\ref{2006C}).
By assumption, the right hand side of (\ref{charl2}) with $\mu_{11}=1$ belongs to $L^{2-\varepsilon}$. It remains only to show that the function
\begin{equation}\label{2006D}
g(k):= \int_{\mathbb C} dS {e^{i \Re(k\overline{z})}} Q_{21}(z) \int_{\mathbb C} d{S_1} \int_{\mathbb C} dS_2 \frac{e^{i \Re(k\overline{z}_1)}}{\overline{z}-\overline{z_1}} \overline{Q}_{12}(z_1) \frac{e^{-i \Re(k\overline{z}_2)}}{z_1-z_2} {Q_{21}}(z_2) \overline{\mu}_{11}(z_2,k)
\end{equation}
belongs to $L^s_k(\mathbb R^2\backslash D), ~ s>{\frac{p}{p-2}}$.

Denote
$$
m(z_1,k) = \int_{\mathbb C}  {e^{i \Re(k\overline{z})}} Q_{21}(z)  \frac{dS}{\overline{z}-\overline{z_1}}.
$$
Since $Q_{21}\in L^p$ and $\frac{1}{z-z_1}\in L^{2-\delta}, ~\delta>0$, the Holder inequality implies that $\frac{Q_{21}(z)}{z-z_1} \in L^{q'}_z$ with arbitrary $q'\in (1,\frac{2p}{2+p})$. Obviously, $1<q'<2$ when $p>4$.  The latter allows one to apply the Hausdorf-Young inequality, which implies that the Fourier transform $m(z_1,\cdot)$ of the function $\frac{Q_{21}(z)}{z-z_1}$ belongs to $L^q_k, ~q> \frac{2p}{p-2}$, uniformly in $z_1\in \mathbb C$ .

We split $\mu_{11}$ in (\ref{2006D}) into two terms: $\mu_{11} = 1+ (\mu_{11}-1)$. Respectively, let $g=g_1+g_2$.  Function $g_1$ (where $\mu_{11}=1$) can be estimated by $C\!\int \!dS_1 |Q_{12}(z_1)m(z_1,k)m(z_1,-k)|$, and therefore, $g_1 \in L^s_k(\mathbb R^2\backslash D), ~ s>{\frac{p}{p-2}}$, due to the Minkowski inequality (written in the integral form). It remains to show that $g_2 \in  L^s_k(\mathbb R^2\backslash D), ~ s>{\frac{p}{p-2}}$. First we estimate the interior part $g_{21}$ of the integral $g_2$:
$$
g_{21}(k,z_1)= \int_{\mathbb C} dS_2 \frac{e^{-i \Re(k\overline{z}_2)}}{z_1-z_2} {Q_{21}}(z_2) (\overline{\mu}_{11}(z_2,k)-1).
$$
Recall that
\begin{equation}\label{1706E}
\sup_z \|\mu(z,\cdot)-I\|_{L^q_k(\mathbb R^2 \backslash D)} \leq C, \quad q>\frac{2p}{p-2}.
\end{equation}
This estimate can be found in \cite[see Th.2.3 and the discussion about the condition $Q=Q^*$ in the proof]{bu}.
Moreover,
\begin{equation}\label{charl4}
\left \| \frac{e^{i \Re(-k\overline{z}_2)}}{z_1-z_2} {Q_{21}}(z_2) \right \|_{L^1_{z_2}(\mathbb R^2)} \leq C
 \end{equation}
  uniformly in $z_1$ due to the Holder inequality and the compactness of the support of  $Q_{21}$. Thus from the integral form of Minkowski's inequality it follows that
\begin{eqnarray}
\|g_{21}\|^q_{L^q_k}  = \left \|\int_{\mathbb C} dS_2 \frac{e^{i \Re(-k\overline{z}_2)}}{z_1-z_2} {Q_{21}}(z_2) (\overline{\mu}_{11}(z_2,k)-1) \right\|_{L^q_k} \leq  \\
\int_{\mathbb C} dS_2 \left | \frac{e^{i \Re(-k\overline{z}_2)}}{z_1-z_2} {Q_{21}}(z_2) \right | \|  \overline{\mu}_{11}(z_2,k)-1\|_{L^q_k} \leq C \|  \overline{\mu}_{11}\overline{}(z_2,k)-1\|_{L^q_k} \leq C_1.
\end{eqnarray}
Function $g_2$ can be estimated by $C\!\int \!dS_1 |m(z_1,k)Q_{12}(z_1)g_{21}(z_1,k)|$.
Finally, applying the integral form of Minkowski's inequality  to the integral in $z_1$, we get that $g_2 \in L^s_k(\mathbb R^2\backslash D), ~ s>{\frac{p}{p-2}}$. Hence, $h_{21}\in L^s_k(\mathbb R^2\backslash D), ~ s>{\frac{p}{p-2}}$. The same inclusion for $h_{12}$ can be proved similarly.

\qed

\begin{lemma}\label{ml2}
For each $z \in \mathbb C$, we have $T_z I \in L^{\widetilde{s}}(\mathbb R^2)$ for each $\widetilde{s}>\max\left (\frac{2p}{p-4},\frac{4}{\varepsilon}\!-\!2\right )$.
\end{lemma}
\noindent
{\bf Proof.} Note that $1<s<2$ for $s$ defined in Lemma \ref{lonE}. Thus Lemma \ref{lonE} and the  Hardy-Littlewood-Sobolev inequality imply that the first term in the right-hand side of (\ref{glavOperator}) with $\phi=I$ belongs to $L^{\widetilde{s}}(\mathbb R^2)$ for each
$\widetilde{s} =\frac{2s}{2-s}$. Thus the statement of the lemma holds for the first term in  (\ref{glavOperator}). The second term in  (\ref{glavOperator}) is continuous in $\mathbb C \backslash \partial D$ with continuous limits at $\partial D$ and has order $1/k$ at infinity. Hence, it also belongs to $L^{\widetilde{s}}(\mathbb R^2)$.

\qed

\begin{lemma}\label{ml1}
The operator $T_z$ is compact in $\mathcal H^s= L^s(\mathbb R^2) \cap C(D)$ for each $s>2$ and depends continuously on $z$.
\end{lemma}
\noindent
{\bf Proof.}
For each function $g(\cdot)$ in $L^2(\mathbb R^2)$, the operator  $\overline{\partial}^{\!~-1}_k (g(k)\cdot)$ is compact on $L^s(\mathbb R^2)$ for all $s>2$ (see, e.g., \cite[Lemma 5.3]{music}) and the following estimate holds (see the same lemma)
\begin{equation}\label{inve}
\|\overline{\partial}^{\!~-1}_k (g(k)\cdot)\|_{L^s(\mathbb R^2)}\leq C \|g(k)\|_{L^2(\mathbb R^2)}.
\end{equation}
From this fact and  Lemma \ref{lonE}, it follows that the operator $T_z^{(1)}$ defined by the first term in the right-hand side of (\ref{glavOperator}) is compact in $L^s(\mathbb R^2), s>2,$ and depends continuously on $z$. While the compactness follows immediately from the references above, in order to prove the continuity in $z$, one needs to
split the operator in two terms $T_z^{(1)}=T_z^{(1,1)}+T_z^{(1,2)}$  by introducing factors $\alpha(\varsigma)$ and $1-\alpha(\varsigma)$ in the integrand in (\ref{glavOperator}), where $\alpha$ is the characteristic function of the region $|\varsigma|>R$. For each $\varepsilon>0$, one can choose $R=R(\varepsilon)$ so large that $\|T_z^{(1,1)}\|<\varepsilon$ for all $z$. This can be done due to (\ref{inve}) and Lemma \ref{lonE}. The continuity of $T_z^{(1,2)}$ follows from  (\ref{inve}) due to the uniform continuity in $z$ of the exponents $e^{i(\overline{\varsigma}z+\overline{z}{\varsigma})/2}$  when $|\varsigma|<R$. Thus operator $T_z^{(1)}$ in $L^s(\mathbb R^2), s>2,$ depends continuously on $z$.

To prove compactness of $T_z^{(1)}$ in $\mathcal H^s$,  we need the following inequality (\cite[Th.1.22]{vekua}): if $f\in L^r(\mathbb R^2)\cap L^q (\mathbb R^2)$ for $1<r<2<q,$ then
\begin{equation}\label{setP1}
\|\overline{\partial}^{\!~-1}f\|_{L^\infty} \leq c_{r,q}(\|f\|_{L^r}+\|f\|_{L^q}).
\end{equation}

Since functions $\phi$ from the domain of operator  $T_z^{(1)}$  belong to $L^s(\mathbb R^2 \backslash D)$ and $h_{12}(\varsigma,\varsigma)$ is smooth in $\overline{\mathbb C \backslash D}$ and belongs to $L^2(\mathbb R^2 \backslash D)$, the product $h_{12}(\varsigma,\varsigma)\phi(\varsigma)$ belongs to $L^r(\mathbb R^2 \backslash D) \cap L^q(\mathbb R^2 \backslash D)$ for some $r,q$ such that $1<r<2<q.$  In fact, Holder's inequality implies that one can take $r=1+\frac{s-2}{s+2},~ q=s$. Thus from (\ref{setP1}) it follows that $\|T_z^{(1)} \phi \|_{L^\infty(\mathbb R^2 )} \leq \|\phi\|_{L^s(\mathbb R^2\backslash D)}$.
Since the range of $T_z^{(1)}$ consists of functions that are holomorphic in $D$, the boundedness of the  set $\left \{T_z^{(1)} \phi, \|\phi\|_{\mathcal H^s}=1 \right \}$ in $C(D)$ implies its compactness in $C(D)$. Hence, the operator $T_z^{(1)}$ is compact in $\mathcal H^s$. Its continuity in $z$ can be proved similarly to continuity in $L^s(\mathbb R^2).$

Let us show the compactness and the continuity in $z$ of the second term  $T_z^{(2)}$ in the right-hand side of (\ref{glavOperator}). We write $T_z^{(2)}$ in the form $T_z^{(2)}=I_1I_2R$, where $R:\mathcal H^s\to C(\partial D)$ is a bounded operator that maps $\phi\in \mathcal H^s$ (recall that $\phi$ belongs to $C(D)$) into its restriction $\phi^-$ on $\partial D$, $I_2 ~:~ C(\partial D) \rightarrow C^{\alpha}(\partial D)$ is defined by the interior integral in the expression for $T_z^{(2)}$, and operator $I_1:C^{\alpha}(\partial D)\to \mathcal H^s$ is defined by the exterior integral in the expression for $T_z^{(2)}$.  Here $C^{\alpha}(\partial D)$ is the Holder space and  $\alpha$ is an arbitrary number in $(0,1/2)$.  The integral kernel of operator $I_2$ has a logarithmic singularity at $\varsigma=\varsigma'$ (due to the presence of the term $W(\varsigma,\varsigma')$). Thus operator $I_2$ is a PDO operator of order $-1$ and therefore $I_2$ is a bounded operator from $C(\partial D)$ into the  Sobolev space $H^1(\partial D)$. Thus it is compact as an operator from $C(\partial D)$ to $C^\alpha(\partial D), \alpha \in (0,1/2)$, due to the Sobolev imbedding theorem. Thus the compactness of  $T_z^{(2)}$ will be proved as soon as we show that $I_1$ is bounded.

For each $\phi \in C^\alpha(\partial D)$, the function $I_1\phi$ is analytic outside of $\partial D$ and vanishes at infinity. Due to the Sokhotski-Plemelj theorem, the limiting values $(I_1\phi)_\pm$ of $(I_1\phi)$ on $\partial D$ from inside and outside of $D$, respectively, are equal to $\frac{\pm\phi}{2}+P.V.\frac{1}{2\pi i} \int_{\partial D} \frac{\phi(\varsigma)d\varsigma}{\varsigma - \lambda}$. Thus
\[
\max_{\partial D}|(I_1\phi)_\pm|\leq C\|\phi\|_{C^\alpha(\partial D)}.
\]
From the maximum principle for analytic functions it follows that the same estimate is valid for the function $I_1\phi$ on the whole plane. Taking also into account that $I_2\phi$ has order $1/k$ at infinity, we obtain that $|I_1 \phi| \leq \frac{C}{1+|k|}\|\phi\|_{C^\alpha(\partial D)}$, i.e.,  operator $I_1$ is bounded. Hence operator $T_z^{(2)}$ is compact.

Obviously, operator $I_2$ depends continuously on $z$, and operators $R$ and $I_1$ do not depend on $z$, i.e., $T_z^{(2)}$ is continuous in $z$.

\qed

\section{Proof of Theorem \ref{mthm} }

Consider the Lippmann-Schwinger equation (\ref{2006A}) for $\mu=\mu_a,~ a>0,$ where the potential $Q$ is replaced by $aQ$. Writing this equation for each component of matrix $\mu_a$ separately, we obtain (compare to (\ref{2006C}))
\begin{eqnarray}\label{0210C}
\overline{\mu}_{a,11}=1+
\frac{a^2}{\pi^2} \int_{\mathbb C} dS_1 \int_{\mathbb C} dS_2 \frac{e^{i \Re(k\overline{z}_1)}}{\overline{z}-\overline{z_1}} \overline{Q}_{12}(z_1) \frac{e^{i \Re(-k\overline{z}_2)}}{z_1-z_2} {Q_{21}}(z_2) \overline{\mu}_{a,11}(z_2,k), \\\label{0210C1}
\mu_{21,a}= a\mathcal L_{k} [Q_{21}(z') \overline{\mu}_{a,11}(z',k)].
\end{eqnarray}
Similar equations hold for $\mu_{a,12},\mu_{a,22}$.
We define $\mu_a$ for complex $a$ via this Lippmann-Schwinger equation.
Let $h_a(\varsigma,k), \varsigma \in \mathbb C, k \in \mathbb C \backslash D,$ be defined by (\ref{14Abr1}) with $\mu_a$ in the integrand instead of $\mu$.
\begin{lemma}\label{lemma710A} Constant $A$ in (\ref{set28A}) can be chosen so large that all the exceptional points for potentials $aQ, ~0\leq a\leq 1,$ are located in the disc $|k|<A-1$. Then point $k_0$ (see (\ref{set28A}), (\ref{eq1})) can be chosen independently of $a$ and the
matrix $h_a(\varsigma,k), \varsigma \in \mathbb C, k \in \mathbb C \backslash D,$ is analytic in $a\in(0,1]$. Moreover, the entries of the derivative  $\frac{\partial h^o_a(k,k)}{\partial a}$ belong to $L^s_k(\mathbb R^2 \backslash D)$ for each $s>\max\left (\frac{p}{p-2},2-\varepsilon \right )$.
\end{lemma}
\noindent
{\bf Proof.}
The first statement follows from the last step in the proof of \cite[Lemma C]{bu}, where it is shown that the norm in $L^\infty_z(L^q_k(\mathbb C \backslash D)), ~q>2p/(p-2),$ of the operator $M$ defined by
$$
(Mv)(z,k) = \int _{\mathbb C}dS_1 \int_{\mathbb C} dS_2 \frac{e^{i \Re(k\overline{z}_1)}}{\overline{z}-\overline{z_1}} \overline{Q}_{12}(z_1) \frac{e^{i \Re(-k\overline{z}_2)}}{z_1-z_2} {Q_{21}}(z_2) \overline{v}(z_2,k)
$$
vanishes when $A \rightarrow \infty$.

From  (\ref{14Abr1}) it follows that $h_a(\cdot,\cdot)$ depends on $a$ only because $\mu=\mu_a(z,k)$ depends on $a$. The choice of $A$ guarantees that the solutions of (\ref{0210C}), (\ref{0210C1}) are analytic in $a\in (0,1]$. Thus $h_a(\cdot,\cdot)$ is analytic in $a\in (0,1]$. Point $k_0$ can be chosen in the region $A-1<|k|<A$. It remains to show that \begin{equation}\label{ink}
\frac{\partial h^o_a(k,k)}{\partial a}\in L^s_k(\mathbb R^2 \backslash D).
\end{equation}

 Differentiation of (\ref{0210C}) in $a$ implies that
\begin{eqnarray}
\frac{\partial \overline{\mu}_{a,11}}{\partial a}-
\frac{a^2}{\pi^2} \int dS_1 \int dS_2 \frac{e^{i \Re(k\overline{z}_1)}}{z-z_1} \overline{Q}_{12}(z_1) \frac{e^{i \Re(-k\overline{z}_2)}}{z_1-z_2} {Q_{21}}(z_2) \frac{\partial \overline{\mu}_{a,11}}{\partial a}(z_2,k) \nonumber \\  \label{0211A} = \frac{2a}{\pi^2} \int dS_1 \int dS_2 \frac{e^{i \Re(k\overline{z}_1)}}{z-z_1} \overline{Q}_{12}(z_1) \frac{e^{i \Re(-k\overline{z}_2)}}{z_1-z_2} {Q_{21}}(z_2)\overline{\mu}_{a,11}(z_2,k).
\end{eqnarray}
It was proved in Lemma \ref{lonE} that  $\mu_{11}-1\in L^\infty_z(L^q_k(\mathbb C \backslash D)), ~q>2p/(p-2)$. Thus the second term in the right-hand side of (\ref{2006C}) belongs to $L^\infty_z(L^q_k(\mathbb C \backslash D)), q>2p/(p-2)$. Lemma \ref{lonE} remains valid when $Q$ is replaced by $aQ$. Hence, the right-hand side of (\ref{0211A}) belongs to $L^\infty_z(L^q_k(\mathbb C \backslash D)), ~q>2p/(p-2)$.
From here and the invertibility of $I-a^2M$ in $L^\infty_z(L^q_k(\mathbb C \backslash D))$ we get that $\frac{\partial  \mu_a}{\partial a}\in L^\infty_z(L^q_k(\mathbb C \backslash D)), ~q>2p/(p-2)$. Now in order to prove (\ref{ink}), one can repeat the arguments from the proof of Lemma \ref{lonE} that were used to estimate $g_2$. The function $\mu_{11}-1$ in these arguments must be replaced by $\frac{\partial  \mu_{11,a}}{\partial a}$.

\qed
\\
{\bf Proof of Theorem \ref{mthm}.}
The first statement of the theorem was proved in Lemma \ref{ml1}. Since $T_z$ is continuous in $z$, the invertibility of $T_z$ at $z=z_0$ implies its invertibility when $|z-z_0|\ll 1$. Thus the second statement of the theorem will be proved if we show
that the set of potentials $Q\in S_{\varepsilon,p}$ (with the support in a fixed $\mathcal O$), for which $I+T_{z_0}$ with a fixed $z_0$ is invertible, is generic, i.e., this set is open and everywhere dense. This set of potentials is open since $T_{z_0}$ depends continuously on $Q$. In order to see that this set of potentials is dense, we note that operator $T_{z_0}$ is analytic in $a\in (0,1]$ due to Lemma \ref{lemma710A}. Obviously, its norm goes to zero as $a\to+0$. Thus operator $I+T_{z_0}$ with $Q$ replaced by $\alpha Q$ is invertible for all $a\in (0,1]$ except at most a finite number of values of $a$. The second statement of the theorem is proved. The third one follows immediately from Lemma \ref{0210L1} and the uniqueness of the solution of (\ref{inteq}).

\qed

\section{Calculation of scattering data via the D-t-N map.}
In this section we discuss the relation between ICP (reconstruction of $\gamma$ via the Dirichlet-to-Neumann map for equation (\ref{set27A}) in $\mathcal O$) and the inverse Dirac scattering problem. In fact, we consider the matrix Lippman-Scwhwinger equation (\ref{19JanA}) instead of the Dirac equation (\ref{fir}) or (\ref{firbc}). The potential $Q$ in (\ref{19JanA}) and $\gamma$ are related via (\ref{char1bc}) and (\ref{char1}), and therefore a reconstruction of $Q$ implies the reconstruction of $\gamma$.  Since we know how to solve the inverse Dirac scattering problem (see Remark 1 after Theorem \ref{mthm}), it remains only to find the Dirac scattering data via the D-t-N map $\Lambda_\gamma$ for equation (\ref{set27A}). Moreover, since the potential $Q$ vanishes outside of $\mathcal O$, the scattering data for the solution of (\ref{19JanA}) can be obtained by simple integration of the Dirichlet data of the same solution $\psi$, see formula (\ref{2106A}). So our aim in this section is to show how the Dirichlet data $\psi|_{\partial\mathcal O}$ for solutions $\psi(z,k)$ of (\ref{19JanA}) with non-exceptional $k$ can be evaluated via the D-t-N map for equation (\ref{set27A}) in $\mathcal O$.

Consider the Faddeev solutions $U_1,U_2$ of equation (\ref{set27A}) in $ \mathbb R^2$, which are determined by the following asymptotic behavior at infinity:
\begin{eqnarray}\label{62}
\frac{i\overline{k}}{2}U_1e^{-\frac{i\overline{k}z}{2}} -1 \rightarrow 0, \quad z \rightarrow \infty, \\\label{63}
\frac{-ik}{2}U_2e^{\frac{ik\overline{z}}{2}} -1 \rightarrow 0, \quad z \rightarrow \infty.
\end{eqnarray}

We will show existence of these solutions when $k$ is not an exceptional point for equation (\ref{19JanA}).
The traces of these solutions on $\partial\mathcal O$ can be found as follows:
\begin{eqnarray}\label{U1}
U_1|_{\partial \mathcal O} = \frac{2}{i\overline{k}}(I+S_{\frac{\overline{k}}{2}}(\Lambda_\gamma-\Lambda_1))^{-1}e^{\frac{iz\overline{k}}{2}} , \\\label{U2}
U_2|_{\partial \mathcal O} = \mathcal C \left [ \frac{2}{i\overline{k}}(I+S_{\frac{\overline{k}}{2}}(\Lambda_{\overline \gamma}-\Lambda_1))^{-1}e^{\frac{iz\overline{k}}{2}} \right ].
\end{eqnarray}
Here $\Lambda_\gamma$ is the D-t-N map for equation (\ref{set27A}) in $\mathcal O$,  $\Lambda_1$ is the same map when $\gamma=1$, and $S_k$ is the single layer operator on the boundary with zero energy Faddeev's Green function:
\[S_k\sigma(z)=\int_{\partial \mathcal O}\mathcal G_k(z-z')\sigma(z')dl_{z'},~~z\in \partial \mathcal O,
\]
where $dl$ is the element of the length and
\[
\mathcal G_k(z)=\frac{1}{4\pi^2} e^{ikx-ky} \int_{\mathbb R^2} \frac{e^{i(\xi_1x+\xi_2 y)}}{|\xi|^2+2k\xi}d\xi_1 d\xi_2, \quad \xi=\xi_1+i\xi_2.
\]
Formulas (\ref{U1}), (\ref{U2}) can be found in \cite[formula (5.18)]{nk}, \cite[formula (26)]{lvf} in the case of the Schr\"{o}dinger operator, but the proofs there can be carried over without any changes to the case of the conductivity equation~(\ref{set27A}).

A point $k\in \mathbb C$ is called {\it exceptional for the Faddeev problem} if equation (\ref{set27A}) in $ \mathbb R^2$ has a nontrivial solution $u$ such that
\begin{equation}\label{uuu}
ue^{-\frac{i\overline{k}z}{2}}  \rightarrow 0 \quad {\rm as} \quad z \rightarrow \infty.
\end{equation}
Obviously, the latter condition is equivalent to $ue^{\frac{ik\overline{z}}{2}}  \rightarrow 0,~ z \rightarrow \infty,$ and corresponds to (\ref{62}), (\ref{63}) with $-1$ in the left-hand sides dropped.

Let $\kappa(z)= \nu_1(z)+ i \nu_2(z)$,  where $\nu=\nu(z)=(\nu_1,\nu_2)$ is the unit outward normal at a point $z \in \partial \mathcal O$, and let
$\partial_s$ be the operator of the tangential (counter-clockwise) derivative on $\partial \mathcal O$.
\begin{theorem}
1) The set $\mathcal E\subset \mathbb C$ of exceptional points for the Dirac problem coincides with the set of exceptional points for the Faddeev problem. The formula
\[
\psi=(\gamma^{1/2}\partial u,\gamma^{1/2}\partial \overline{u})^t, \quad k\in \mathcal E,
\]
establishes a one-to-one correspondence between vector solutions $\psi $ of homogeneous equation (\ref{19JanA}) and solutions $u$ of (\ref{set27A}) in $ \mathbb R^2$ that satisfy (\ref{uuu}).

2) The formula
\begin{equation}\label{psu}
\psi(z,k)= \left ( \begin{array}{cc}
\gamma^{1/2}\partial U_1 &   \gamma^{1/2}\partial U_2\\
\overline{\gamma^{1/2}}\partial \overline{U_1} &\overline{\gamma^{1/2}}\partial \overline{U_2}
\end{array} \right ), \quad k\in \mathbb C\backslash \mathcal E,
\end{equation}
establishes a one-to-one correspondence between the Faddeev solutions $(U_1,U_2)$ and the scattering solutions $\psi$ of the Dirac equation (\ref{fir}) that satisfy (\ref{lim}).

3) The Dirichlet data $\psi|_{\partial \mathcal O}$ of the scattering solutions of the Dirac problem can be found as follows
\begin{equation}\label{1112A}
\left (
\begin{array}{cc}
\psi_{11} &   \psi_{12}\\
\overline{\psi_{21}} &   \overline{\psi_{22}}
\end{array} \right )= \frac{1}{2}
\left ( \!\begin{array}{cc} \overline{\kappa} & -i\overline{\kappa} \\ \kappa  & i\kappa \end{array} \!\right )
\left ( \!\!\!\begin{array}{cc} \Lambda_\gamma  U_1 & \Lambda_\gamma  U_2  \\ \partial_s U_1 & \partial_s U_2\end{array} \!\!\!\right ),\quad z\in \partial \mathcal O, \quad k\notin \mathcal E.
\end{equation}
\end{theorem}
\noindent
{\bf Remarks.} 1) In order to apply (\ref{1112A}), one needs only to know $\Lambda_\gamma$ and to be able to evaluate the right-hand sides in  (\ref{U1}), (\ref{U2}).

2) Recall that the problems (\ref{fir})-(\ref{lim}) and (\ref{19JanA}) are equivalent.
\\
{\bf Proof.} Let us prove the second statement. Let $(U_1,U_2)$ be the Faddeev solutions. It was shown in \cite{bu}, \cite{fr} (and can be easily verified) that the matrix function (\ref{psu}) satisfies the Dirac equation (\ref{fir}). Since formulas (\ref{62}), (\ref{63}) admit differentiation, one can also check that matrix (\ref{psu}) satisfies (\ref{lim}). (Note that $\partial \overline{U_1}e^{\frac{ik\overline{z}}{2}} \rightarrow 0, ~ z \rightarrow \infty,$ is equivalent to $\partial \overline{U_1}e^{-\frac{i\overline{k}z}{2}}, ~ z \rightarrow \infty.$ This simple fact is needed in order to establish the asymptotic behavior for the off-diagonal terms in (\ref{psu}).)

Conversely, let $\psi=(\psi_1,\psi_2)^t$ be the first column of the solution of (\ref{fir}), (\ref{lim}). Let $\phi_1=\psi_1$ and $\phi_2=\overline{\psi}_2$. Then $\phi=(\phi_1,\phi_2)^t$
 is a solution of (\ref{firbc}). From (\ref{firbc}) and (\ref{char1bc}) it follows that the compatibility condition
$$
\overline{\partial} (\gamma^{-1/2} \phi_1) = {\partial} (\gamma^{-1/2} \phi_2)
$$
holds. Then the Poincare lemma implies the existence of such a $U_1$ that
\begin{equation}\label{lll}
 \left(\!\!\!\begin{array}{c} \phi_1 \\ \phi_2 \end{array} \!\!\!\right ) =
\gamma^{-1/2}\left ( \!\!\!\begin{array}{c} \partial U_1 \\ \overline{\partial} U_1\end{array} \!\!\!\right ).
\end{equation}
By applying operator $\overline{\partial}$ to the first components of vector equation (\ref{lll}) (or applying $\partial$ to the second components) and using (\ref{firbc}), (\ref{char1bc}), one can proof that $U_1$ satisfies equation (\ref{set27A}) in $ \mathbb R^2$. Moreover, $U_1$ can be represented in the form of a contour integral involving $\psi$, and asymptotics  (\ref{lim}) of $\psi$ admits differentiation. Using integration by parts, one can show that this representation of $U_1$ implies (\ref{62}). This justifies the equality of the first columns in (\ref{psu}). Function $U_2$ can be constructed similarly. The second statement of the theorem is proved. The proof of the first statement is no different.
 Relation  (\ref{1112A}) follows from (\ref{psu}) if one applies the complex conjugation to the second rows in (\ref{psu}) and expresses the vector $(\partial,\overline{\partial})^t$ there  via $(\partial_\nu,\partial_s)^t$.

  \qed

{\bf Acknowledgments.}  The authors are thankful to Eemeli Bl\aa sten, Boaz Haberman, Roland Griesmaier, and Lassi P$\rm{\ddot{a}}$iv$\!\rm{\ddot{a}}$rinta for useful discussions.


\begin{thebibliography}{102}

\bibitem{afr} Astala, K., Faraco, D., Rogers, K. M. (2016), Unbounded potential recovery in the plane. Annales scientifiques de l’École Normale Supérieure.
49, 1023–1047.

\bibitem{ap}
Astala K., P$\ddot{a}$iv$\ddot{a}$rinta L. (2006). Calderón's inverse conductivity problem in the plane, Annals of Mathematics, 265-299.


  \bibitem{bc1}
Beals R.,   Coifman R.R., (1985). Multidimensional inverse scatterings and
nonlinear partial differential equations. In F. Treves, editor, Pseudodifferential operators and applications, volume 43 of Proceedings of symposia
in pure mathematics, Amer. Math. Soc., 45-70.


  \bibitem{bc2}
Beals R., Coifman R.R. (1988). The spectral problem for the Davey-
Stewartson and Ishimori hierarchies. In Nonlinear evolution equations:
Integrability and spectral methods, 15-23. Manchester University
Press.

\bibitem{blasten}
Bl\aa sten, E., Imanuvilov, O. Y., Yamamoto, M. (2015). Stability and uniqueness for a two-dimensional inverse boundary value problem for less regular potentials,  Inverse Problems and Imaging, 9(3).


\bibitem{borcea} Borcea L., (2002). Electrical impedance tomography, Inverse Problems 18, 99–136.

  \bibitem{bu}
Brown R. M., Uhlmann G. A. (1997). Uniqueness in the inverse conductivity problem for nonsmooth conductivities in two dimensions, Communications in partial differential equations, 22(5-6), 1009-1027.



\bibitem{bukh} Bukhgeim A. L. (2008). Recovering a potential from Cauchy data in the
two-dimensional case, J. Inverse Ill-Posed Probl., 16(1), 19-33


\bibitem{fr}
 Francini E., (2000). Recovering a complex coefficient in a planar domain from
Dirichlet-to-Neumann map, Inverse Problems, 16, 107–119.


\bibitem{hs}
Hamilton, S. J.,  Siltanen, S. (2014). Nonlinear inversion from partial EIT data: computational experiments, Inverse problems and applications, 615, 105-129.

\bibitem{knud} Knudsen, K., Tamasan, A. (2005). Reconstruction of less regular conductivities in the plane. Communications in Partial Differential Equations, 29(3-4), 361-381.

\bibitem{lvf} Lakshtanov, E.,  Vainberg, B. (2014). Exceptional points in Faddeev scattering problem, to appear in Theor.Math.Phys.

\bibitem{lnv} Lakshtanov E., Novikov R., Vainberg B. (2016). A global Riemann-Hilbert problem for two-dimensional inverse scattering at fixed energy,  	 Rend. Istit. Mat. Univ. Trieste, 48, 1–26.


\bibitem{lnv2} Lakshtanov E., Vainberg B.  Recovery of $L^p$-potential in the plane, in preparation.

\bibitem{music} Music, M. (2014). The nonlinear Fourier transform for two-dimensional subcritical potentials, Inverse Problems and Imaging, 8(4),  11511167.

\bibitem{nachman}
Nachman A. I. (1996). Global uniqueness for a two-dimensional inverse boundary value problem, Annals of Mathematics, 71-96.

  \bibitem{RN2}
  Novikov, R. G. (1992). The Inverse Scattering Problem
on a Fixed Energy Level for the Two-Dimensional Schr\"{o}dinger Operator, J. of Funct. Analysis, 103, 409-463.

\bibitem{nk}
Novikov, R. G.,  Khenkin, G. M. (1987). The $\overline{\partial}-$equation in the multidimensional inverse scattering problem. Russian Mathematical Surveys, 42(3), 109-180.

\bibitem{ns}
Novikov R. G., Santacesaria M., (2011). Global uniqueness and reconstruction for the
multi-channel Gel’fand-Calder´on inverse problem in two dimensions, Bull. Sci. Math.
135, 5, 421–434.

\bibitem{sung1}
Sung, L. Y. (1994). An Inverse Scattering Transform for the Davey-Stewartson-II Equations, I. Journal of mathematical analysis and applications, 183(1), 121-154.





\bibitem{vekua} Vekua I. N. (2014). Generalized analytic functions, International Series of Monographs on Pure and Applied Mathematics, (25), Elsevier.

\end{thebibliography}
\end{document}